\documentclass[AMA,STIX1COL]{WileyNJD-v2}

\usepackage{tabularx,booktabs}
\usepackage{bbold}
\usepackage{color,colortbl}
\usepackage{bbm}
\usepackage{mathtools} 
\usepackage{etoolbox}
\usepackage{dsfont}
\usepackage{booktabs,caption,fixltx2e}
\usepackage{multirow}
\usepackage{graphicx}
\usepackage{tabularx,booktabs}
\usepackage{bbold}
\usepackage{subcaption}
\usepackage{color,colortbl}
\usepackage{bbm}
\usepackage{mathtools} 
\usepackage{etoolbox}
\usepackage{dsfont}
\usepackage{booktabs,caption,fixltx2e}
\usepackage[flushleft]{threeparttable}
\usepackage{multirow}
\usepackage{graphicx}
\usepackage{setspace}
\usepackage[utf8]{inputenc}
\usepackage{enumitem} 

\usepackage{comment}

\hypersetup{
    colorlinks,
    citecolor=black,
    filecolor=black,
    linkcolor=black,
    urlcolor=black
}

\AtBeginEnvironment{tabular}{\singlespacing}

\BeforeBeginEnvironment{verbatim}{\def\baselinestretch{1}}

\renewcommand{\baselinestretch}{1.4}
\def\singlespace{\def\baselinestretch{1}\@normalsize}

\usepackage{bm}

\catcode`@=11

\def\@evenhead{\vbox{\hbox to\textwidth{\footnotesize \hfill
 \hfill } }}
\usepackage{amsmath,graphicx,centernot}

\def\P{{\mbox{\rm\tiny P}}}

\def\L{{\mbox{\rm\tiny L}}}

\articletype{}

\received{}
\revised{}
\accepted{}

\begin{document}

\title{Sample size planning for estimating the global win probability with assurance and precision}

\author[1,2]{Di Shu*}

\author[3,4]{Guangyong Zou}


\address[1]{Department of Biostatistics, Epidemiology and Informatics, University of Pennsylvania Perelman School of Medicine, Philadelphia, PA, USA}

\address[2]{Clinical Futures, Children’s Hospital of Philadelphia, Philadelphia, PA, USA}

\address[3]{Department of Epidemiology and Biostatistics, Schulich School of Medicine \& Dentistry, Western University, London, ON, Canada}

\address[4]{Robarts Research Institute, Western University, London, ON, Canada}

\corres{*Di Shu, Department of Biostatistics, Epidemiology and Informatics, Perelman School of Medicine, University of Pennsylvania, 423 Guardian Drive Philadelphia, PA 19104-6021, USA\\
\email{shudi1991@gmail.com}}

\abstract{Most clinical trials conducted in drug development contain multiple endpoints in order to collectively assess the intended effects of the drug on various disease characteristics. Focusing on the estimation of the global win probability, defined as the average win probability (WinP) across endpoints that a treated participant would have a better outcome than a control participant, we propose a closed-form sample size formula incorporating pre-specified precision and assurance, with precision denoted by the lower limit of confidence interval and assurance denoted by the probability of achieving that lower limit. We make use of the equivalence of the WinP and the area under the receiver operating characteristic curve (AUC) and adapt a formula originally developed for the difference between two AUCs to handle the global WinP. Unequal variance is allowed. Simulation results suggest that the method performs very well. We illustrate the proposed formula using a Parkinson’s disease clinical trial design example.}



\maketitle

\section{Introduction}\label{introduction}

Most clinical trials conducted in drug development contain multiple endpoints in order to collectively assess the intended effects of the drug on various disease characteristics.\citep{FDA2022} These endpoints could include assessments of clinical events, symptoms, measures of function and more. For example, in Parkinson’s disease (PD) clinical trials aiming to identify a neuroprotective treatment that can slow or stop PD progression, more than one aspects are needed to capture PD disability, including motor and nonmotor functions, cognition, and drug complications.\citep{Huang2008}

Sample size planning is a key task in trial design. Too big or small a sample size planned can considerably waste  resources and even be unethical.
Significant advancements  have been made to provide methods for sample size estimation in clinical trials with multiple endpoints. \citep{Huang2008,Lafaye2014} We take a different sizing approach focusing on the estimation of the global win probability---the average win probability (WinP) across endpoints that a treated participant would do better than a control participant---with pre-specified precision and assurance, with precision denoted by the lower limit of confidence interval (CI) and assurance denoted by the probability of achieving that lower limit. Our CI approach to sample size planning is consistent with the trend towards emphasizing estimation over testing.\citep{Amrhein2019}

\section{Methods}\label{methods}
\subsection{Estimand and its estimation}\label{est}

Suppose we have $K$ endpoints. For $k=1,\ldots, K$, let $X_{1k}$ and $X_{0k}$ denote the $k$th endpoint of the treated and control arms, respectively. Regarding a higher outcome as better, we define the treatment effect on the $k$th endpoint as 
\begin{equation}
\theta_k=P(X_{1k}>X_{0k})+0.5P(X_{1k}=X_{0k})
\label{equ1}
\end{equation}
for  $k=1,\ldots, K$. Here $\theta_k$ is referred to as the win probability because it represents the probability that a treated participant would have a better outcome than a control participant; a 50:50 win is given when a tie occurs. \citep{Colditz1988,Zou2022,Zou2023cct,Zou2023ps} An overall treatment effect on the $K$ endpoints can be summarized by the global WinP defined as the average of the endpoint-specific WinPs:
\begin{equation}
\theta=\dfrac{1}{K}\sum_{k=1}^K \theta_k .
\label{equ2}
\end{equation}
The global WinP was also considered in cluster randomized trials with multiple endpoints. \citep{Davies2024}

Since the  WinP and the area under the receiver operating characteristic curve (AUC)  are equivalent,\citep{Bamber1975} we can directly apply  the Delong nonparametric method,\citep{Delong1988} originally developed for AUCs and their differences, to obtain endpoint-specific WinP estimates and their variances and covariances. Let $\hat\theta_k$ and $\widehat{var}(\hat\theta_k)$ denote the point and variance estimates for $\theta_k$ in (\ref{equ1}), for $k=1,\ldots, K$. Let $\widehat{\text{cov}}_{ij}$ denote the covariance between $\theta_i$ and $\theta_j$, for $1\le i <j \le K$. Then, the global WinP can be estimated as $\hat\theta=\sum_{k=1}^K \hat\theta_k/K$ with  $\widehat{\mbox{var}}(\hat\theta)=\{\sum_{k=1}^K\widehat{\mbox{var}}(\hat\theta_k)+2\sum_{1\le i<j\le K}\widehat{\mbox{cov}}_{ij}\}/K^2$.
It is natural to conduct inference for the WinP based on the logit transformation, as it is a probability by definition. Specifically, the CI for $\theta$ is given by\begin{equation}\dfrac{\exp(l)}{1+\exp(l)} \quad \text{to} \quad \dfrac{\exp(u)}{1+\exp(u)}
\label{equ3}
\end{equation}
where 
\begin{equation}
l,u=\log\left(\dfrac{\hat\theta}{1-\hat\theta}\right) \pm z_{0.025}\left\{\dfrac{\sqrt{\widehat{var}(\hat\theta)}}{\hat\theta(1-\hat\theta)}\right\} 
\label{equ4}
\end{equation}
and $z_{0.025}$ is the upper $2.5\%$ quantile of $N(0,1)$, the standard normal distribution. This transform performs well for the estimation of single WinPs.\citep{Zou2023cct,Zou2023ps}

\subsection{Sample size formula for estimating the global WinP}\label{proposed}

We aim to compute required sample size such that $P(\hat\theta_{\L}\ge \theta_0)\ge 1-\beta$ for a given $\theta>\theta_0$, where $\hat\theta_{\L}$ denotes the lower confidence limit for $\theta$ and $\theta_0$ denotes a pre-specified lower bound for $\hat\theta_{\L}$. Here $1-\beta$ is referred to as the assurance probability, which describes the confidence in estimating the global WinP with an acceptable precision.\citep{Zou2012,ShuZou2023} While assurance may be translated into the hypothesis testing framework, presenting based on the CI is more cohesive with the goal of estimation.

Shu and Zou\citep{ShuZou2023} proposed a closed-form sample size formula for the  estimation of the difference between two AUCs. Making use of the equivalence of the AUC and WinP, we adapt the formula to handle the global WinP and propose to calculate the total sample size  as
\begin{equation}
n=\left\{\dfrac{z_\beta+z_{\alpha/2}}{\text{logit}(\theta)-\text{logit}(\theta_0)}\right\}^2 \dfrac{f(\theta)}{\theta^2(1-\theta)^2}\dfrac{\pi}{3} 
\label{equ5}
\end{equation}
where $\alpha$ is typically  $5\%$ corresponding to a $95\%$ CI, $z_{\beta}$ is the upper $\beta$ quantile of $N(0,1)$,
\begin{equation}
f(\theta)=\dfrac{1}{K^2}\left\{\sum_{k=1}^K f^{(k)}(\theta_k)+2\sum_{1\le i<j\le K}\rho_{ij}\sqrt{f^{(i)}(\theta_i)}\sqrt{f^{(j)}(\theta_j)} \right\}
\label{equ6}
\end{equation}
and
\begin{equation}
f^{(k)}(\theta_k)=\dfrac{1}{2}[\varphi\{\Phi^{-1}(\theta_k)\}]^2 \left[\dfrac{\{\Phi^{-1}(\theta_k)\}^2}{(1+B_k^2)^2}\left\{r+1+\dfrac{(r+1)B_k^4}{r}\right\}+\dfrac{2(r+1)}{1+B_k^2}+\dfrac{2(r+1)B_k^2}{r(1+B_k^2)}\right] .
\label{equ7}
\end{equation}
Here $\rho_{ij}$ is the correlation between two estimated WinPs from endpoints $i$ and $j$, $r$ is the
sample size ratio of control group to treated group, $B_k$ is
the standard deviation
ratio of control group to treated group for endpoint $k$,  $\Phi^{-1}(\cdot)$ is the inverse of the cumulative distribution function of $N(0,1)$, and $\varphi(\cdot)$ is the probability density function of $N(0,1)$. 

Similar to Shu and Zou,\citep{ShuZou2023} the proposed formula assumes data are normally distributed.  The inflation factor $\pi/3$ in (\ref{equ5}), which is the reciprocal of the nonparametric statistic relative
efficiency,\citep{Lehmann2005} is set to account for the discrepancy between design and analysis stages. Practical sample size formulas  often require certain parametric assumptions when designing a study, whereas nonparametric estimation, if available, may be employed in the analysis stage for robustness considerations.

To ensure integer sample size, we calculate the treated and control group sizes as $\text{ceil}\{n/(r+1)\}$ and $\text{ceil}\{rn/(r+1)\}$, respectively, where $\text{ceil}(\cdot)$ indicates the ceiling function. 

Between-WinP correlations in (\ref{equ6}) can be obtained from pilot data or raw data correlations. \citep{ShuZou2023} Considering it is common to have positively correlated endpoints and that the raw data correlation tends to be slightly larger than the between-WinP correlation, a practical  approach is to use raw data correlations when implementing (\ref{equ6}).

\section{Simulation} \label{simulation}

\subsection{Set-up} \label{setup}
We conducted a simulation study to evaluate the performance of the proposed method. Two criteria were used: i) the empirical assurance  defined as the percentage of the lower confidence limits in 10,000 simulation
runs that exceeded the pre-specified lower bound, and ii) the empirical coverage defined as the percentage of 95\% CIs in 10,000 simulation runs that covered the true global WinP. Satisfactory performance requires  empirical assurance  close to the pre-specified assurance  and empirical coverage close to 95\%.

We considered three endpoints and set their WinPs  to be $\theta_1 = 0.7$, $\theta_2 = 0.65$ and $\theta_3 = 0.6$, resulting in a global WinP of $\theta=0.65$. We specified the assurance  as 80\% or 90\%, the lower bound  as  0.55 or 0.6, the group size ratio as 1 or 2, and the standard deviation ratio as 1 or 2 for each endpoint (and thus let $B$  denote this common ratio). The control group data followed $N(0,1)$ and the treated group data followed  a normal distribution whose mean and standard deviation were determined by the underlying
configuration. A between-endpoint correlation for outcome data was set to 0.75 or 0.15 to indicate high or low degrees of correlation. We used the raw data correlation to approximate the between-WinP correlation when calculating the sample size. Once obtaining the sample size, for each  scenario we simulated 10,000 data sets each having that size. For each simulated data set, the 95\% CI were constructed as described in Section \ref{est}.

\subsection{Results} \label{setup}

Table \ref{tb1}  summarizes simulation results. The empirical assurance  was close to the pre-specified  in all
scenarios considered. For example, the formula predicts that if the three endpoint-specific WinPs are 0.7, 0.65 and 0.6 with a pairwise correlation of $0.75$, a randomized trial of a total sample size of 286 with equal arm size and common variance would provide a
lower limit  to be above 0.55 with 90\% assurance,
comparable with 91.00\% as estimated based on 10,000 simulation runs.
The empirical coverage was very close to 95\% in all cases,  demonstrating reliable performance of the DeLong nonparametric method
when combined with the logit transformation under a total sample size as small as 102.
The required sample size increased with the correlation between endpoints, suggesting a need to properly account for these correlations  in the design stage.

\begin{center}
{\it [insert Table \ref{tb1} here]}
\end{center}

\section{Illustrative example: PD clinical trial design} \label{example}

Now, we use the PD clinical trial design example, previously considered by Huang et al., \citep{Huang2008}  to illustrate the proposed method. Suppose a future phase 3 trial is to  compare creatine with placebo in the following five endpoints: the changes from baseline to  5-year follow-up in the measures of the modified Rankin, symbol digit modalities, Schwab and England ADL scale, PDQ-39, and ambulatory capacity. Converting endpoint-specific treatment effects assumed by Huang et al. to the WinP scale, we obtain $\theta_1=0.593$, $\theta_2=0.556$, $\theta_3=0.551$, $\theta_4=0.544$, $\theta_5=0.553$ and $\theta=0.559$.  

Consider a lower bound of 0.5. Table \ref{tb2} reports the design results under various scenarios. For example, with a pairwise correlation of 0.3 between endpoints, a total of 492 participants are required in a 2:1 randomization trial with common variance such that we have 90\% assurance  of  the lower confidence limit of the global WinP being at least 0.5.

A phase 2 study suggested correlations among the five outcomes do not exceed 0.35. \citep{Huang2008} If similar patterns can be assumed for the new trial, one could specify the correlation as 0.5 to allow for some uncertainty.

\begin{center}
{\it [insert Table \ref{tb2} here]}
\end{center}

\section{Concluding remarks} \label{discussion}
We have proposed a closed-form sample size formula  for achieving  an acceptable lower bound of the CI for the global WinP with  desired assurance. Our method is flexible to allow variance heterogeneity between treatment groups and factors in the discrepancy in variance estimation comparing the design and analysis stages. Simulation results demonstrated satisfactory performance of the proposed formula.  Inappropriately assuming a common variance can lead to under- or over- planned  sample size (e.g., see the last three rows of Table \ref{tb2}).

Our work relies on the normality assumption. This parametric restriction could be mitigated by proper transformation, because the method is applicable to any data that can be normalized with monotone transformation.

\section*{CONFLICT OF INTEREST STATEMENT}
The authors declare no conflicts of interest.

\section*{DATA AVAILABILITY STATEMENT}
Not applicable.

\bibliography{MRref}

\newpage

\begin{table}[!t]
\centering
\tabcolsep12pt
\caption{Performance of the proposed sample size formula for estimating a global WinP assumed to be $\theta = 0.65$ (as an average of three endpoint-specific WinPs $\theta_1 = 0.7$, $\theta_2 = 0.65$ and $\theta_3 = 0.6$) with pre-specified lower limit ($\theta_0$) and assurance
of 80\% or 90\%, under various combinations of between-endpoint correlations, standard deviation ratio of control group to treated group ($B$) and
sample size ratio of control group to treated group ($r$).}
\begin{threeparttable}
\begin{tabularx}{\textwidth}{ccccccccccc}
\toprule
            &        &   &   & \multicolumn{3}{c}{80\% Assurance} &  & \multicolumn{3}{c}{90\% Assurance} \\ \cline{5-7} \cline{9-11} 
Correlation & $\theta_0$ & $B$ & $r$ & $n$        & ECP        & EAP        &  & $n$         & ECP        & EAP       \\ \hline
High        & 0.55   & 1 & 1 & 214      & 95.24      & 81.29      &  & 286       & 94.82      & 91.00        \\
            &        &   & 2 & 240      & 95.10       & 81.70       &  & 321       & 94.70       & 91.17     \\
            &        & 2 & 1 & 216      & 95.20       & 80.48      &  & 290       & 95.16      & 90.33     \\
            &        &   & 2 & 194      & 95.11      & 80.32      &  & 260       & 95.32      & 91.15     \\
            & 0.6    & 1 & 1 & 818      & 95.36      & 81.61      &  & 1096      & 95.04      & 90.88     \\
            &        &   & 2 & 921      & 95.17      & 81.82      &  & 1232      & 94.90       & 91.53     \\
            &        & 2 & 1 & 830      & 94.94      & 79.75      &  & 1110      & 95.20       & 89.80      \\
            &        &   & 2 & 743      & 95.07      & 80.05      &  & 993       & 94.86      & 90.74     \\
Low         & 0.55   & 1 & 1 & 112      & 95.08      & 82.27      &  & 150       & 94.85      & 91.62     \\
            &        &   & 2 & 126      & 94.99      & 81.37      &  & 168       & 95.02      & 91.46     \\
            &        & 2 & 1 & 114      & 95.16      & 80.23      &  & 152       & 94.52      & 90.42     \\
            &        &   & 2 & 102      & 94.95      & 79.87      &  & 135       & 94.92      & 90.47     \\
            & 0.6    & 1 & 1 & 426      & 94.88      & 81.33      &  & 570       & 94.84      & 91.29     \\
            &        &   & 2 & 480      & 95.29      & 81.62      &  & 642       & 95.48      & 90.99     \\
            &        & 2 & 1 & 432      & 94.86      & 79.25      &  & 578       & 94.66      & 89.17     \\
            &        &   & 2 & 387      & 94.89      & 79.71      &  & 518       & 95.45      & 89.67     \\
\bottomrule
\end{tabularx}
 {
$n$, estimated total sample size; ECP, empirical coverage, estimated by percentage of
times that the 2-sided 95\% CIs contain the true value of the global WinP across 10,000
simulated data sets; EAP, empirical assurance, estimated by percentage of times
that the lower limits of 2-sided 95\% CIs exceed the pre-specified lower bound
across 10,000 simulated data sets.
     }
     \end{threeparttable}
\label{tb1}
\end{table}

\clearpage
\newpage

\begin{table}[!t]
\centering
\tabcolsep12.5pt
\caption{Required total sample size for the illustrative PD clinical trial design example obtained using the proposed method, under assumed endpoint-specific WinPs of 0.593, 0.556, 0.551,  0.544 and 0.553 with a global WinP of 0.559, a lower bound $\theta_0$ of 0.5 and various combinations of assurance, between-endpoint correlation ($\rho$),  standard deviation ratio of control group to treated group ($B$) and
sample size ratio of control group to treated group ($r$).}
\begin{tabularx}{\textwidth}{cccccccccc}
\toprule
    &     &  & \multicolumn{3}{c}{80\% assurance} &  & \multicolumn{3}{c}{90\% assurance} \\ \cline{4-6} \cline{8-10} 
$r$   & $B$   &  & $\rho=0.1$    & $\rho=0.3$   & $\rho=0.5$   &  & $\rho=0.1$   & $\rho=0.3$   & $\rho=0.5$   \\ \hline
1   & 0.5 &  & 210        & 328       & 448       &  & 280        & 440       & 598       \\
    & 1   &  & 208        & 328       & 446       &  & 280        & 438       & 598       \\
    & 2   &  & 210        & 328       & 448       &  & 280        & 440       & 598       \\
0.5 & 0.5 &  & 188        & 296       & 402       &  & 252        & 395       & 539       \\
    & 1   &  & 234        & 368       & 501       &  & 314        & 492       & 672       \\
    & 2   &  & 282        & 443       & 603       &  & 378        & 593       & 807       \\ \bottomrule
\end{tabularx}
\label{tb2}
\end{table}

\end{document}